\documentclass[notitlepage,aps,letter,reprint,nofootinbib,superscriptaddress]{revtex4-1}

\usepackage{graphicx}
\usepackage{amssymb}
\usepackage{bbold}
\usepackage{verbatim}

\usepackage{amsmath}

\newcommand{\be}{\begin{eqnarray*}}
\newcommand{\ee}{\end{eqnarray*}}
\newcommand{\beq}{\begin{eqnarray}}
\newcommand{\eeq}{\end{eqnarray}}
\newcommand{\bequ}{\begin{equation}}
\newcommand{\eequ}{\end{equation}}
\newcommand{\ket}[1]{\left|{#1}\right\rangle}
\newcommand{\bra}[1]{\left\langle{#1}\right|}
\newcommand{\zbb}{\mathbb{Z}}
\newcommand{\id}{\mathbb{I}}

\newcommand{\dd}{\mathrm{d}}
\renewcommand{\epsilon}{\varepsilon}
\renewcommand{\vec}[1]{{\bf #1}}

\begin{document}
\setlength{\pdfpagewidth}{8.5in}%
\setlength{\pdfpageheight}{11in}% 

\title{Topology and Broken Symmetry in Floquet Systems}

\author{Fenner Harper}
\affiliation{Mani L. Bhaumik Insitute for Theoretical Physics, Department of Physics and Astronomy, UCLA, Los Angeles CA 90095, USA}
\author{Rahul Roy}
\affiliation{Mani L. Bhaumik Insitute for Theoretical Physics, Department of Physics and Astronomy, UCLA, Los Angeles CA 90095, USA}
\author{Mark S. Rudner}
\affiliation{Center for Quantum Devices and Niels Bohr International Academy,Niels Bohr Institute, University of Copenhagen, 2100 Copenhagen, Denmark}
\author{S. L. Sondhi}
\email{sondhi@princeton.edu}
\affiliation{Department of Physics, Princeton University, Princeton NJ 08544, USA}

%\date{March 2019}
\begin{abstract}
Floquet systems are governed by periodic, time-dependent, Hamiltonians. 
{\it Prima facie} they should absorb energy from the external drives involved in modulating their couplings and heat up to infinite temperature. However this unhappy state of affairs can be avoided in many ways. Instead, as has become clear from much recent work, they can exhibit a variety of nontrivial behavior---some of it impossible in undriven systems. In this review we describe the main ideas and themes of this work: 
novel Floquet drives which exhibit nontrivial topology in single-particle systems, the existence and classification of exotic Floquet drives in interacting systems, and the attendant notion of many-body Floquet phases and arguments for their stability to heating.
\end{abstract}

\maketitle

%%%%%%%%%%%%%%%%%%%%%%%%%%%%%%%%%%%%%%%%%%%%%%%%%%%%
\section{Introduction}
%%%%%%%%%%%%%%%%%%%%%%%%%%%%%%%%%%%%%%%%%%%%%%%%%%%%
Fundamentally, quantum mechanics in experimental settings is concerned with the time evolution of a system coupled with its environment including the measuring apparatus. As such, it is the unitary time-evolution operator that should play a central role in their analysis. However, for a very large class of systems, including the field theories of particle physics outside of cosmological settings, one can reduce the core of the problem to understanding the behavior of a static system with a time-independent Hamiltonian. So while it remains true that the unitary time-evolution operator is the object of ultimate interest, in practice the Hamiltonian $H$ is the star of the show. We learn vast amounts by finding its eigensystem: specifically, the eigenstates give rise to special, stationary, 
solutions of the Schr\"odinger equation that form a basis for general time evolution. For sufficiently ergodic many-body systems the Hamiltonian obeys the eigenstate thermalization hypothesis (ETH) and, consequently, late-time states obtained from fairly general starting states are effectively described by the Gibbs state $\rho_G \propto e^{-\beta H}$ determined by $H$.

Recently much interest and energy has been focused on a class of quantum systems where this simplification is not possible. These are periodically driven or Floquet systems, which are governed by time-dependent Hamiltonians $H(t)$ which repeat with a fixed period $T$: $H(t) \equiv H(t+T)$. As the Hamiltonian is now time dependent, it is now necessary to work with the unitary operator, which brings an unfamiliar set of challenges. However, thanks to the periodicity there is still some structure in the time domain that allows for mathematical simplification and, more importantly, for physical complexity that would not be possible in its absence. This complexity encompasses nontrivial topology in single-particle Floquet physics, the phenomenon of many body Floquet localization, and the existence of localized interacting phases based either on broken symmetries or on nontrivial topology which have no analog in undriven systems. In this brief review we present the highlights of this body of work in this order. 

Of course, the study of periodically driven systems has a long and varied history, and in this review we do not claim to give a comprehensive record. The interested reader may also wish to consult some of the existing pedagogical works in the Floquet literature, notably reviews which discuss Floquet topological insulators \cite{Cayssol2013}, band structure stabilization and engineering \cite{BukovReview,Oka2019}, periodically driven optical lattices \cite{Holthaus2015,Eckardt2017}, as well as some earlier approaches to the study of driven systems \cite{Shirley65, Sambe1973,dittrich1998,Holthaus1996,Kohler2005}.

Before proceeding we should offer one clarification. The most general setting beyond closed systems with time-independent Hamiltonians is that of open quantum systems (see Ref.~\citenum{breuer2007} for a review). While their most general description is in terms of a unitary acting on the system and its environment, much effort has gone into developing descriptions in which the environment is integrated out in favor of a more complex, non-Hamiltonian, description of the system itself. Floquet systems are a special subclass in which there is still a Hamiltonian for the system and the effect of the environment (the ``drive'') is encompassed entirely in a periodic modulation of various couplings present in it. The system exchanges energy with the environment but does not get entangled with it.

%%%%%%%%%%%%%%%%%%%%%%%%%%%%%%%%%%%%%%%%%%%%%%%%%%%%
\section{Basic Formalism}
\label{sec:basic_formalism}

For a Floquet system we need to solve the the time-dependent Schr\"odinger equation ($\hbar=1$) 
\beq
\label{eq:unitaryevolution}i\frac{d}{dt} U(t,t_0)= H(t) U(t,t_0),
\eeq
where $U(t,t_0)$ is the unitary time-evolution operator that relates states at time $t_0$ to states at time $t$. A given Floquet ``drive'' is specified by a periodic operator-valued function of time, $H(t+T) \equiv H(t)$. Thanks 
to the periodicity of the Hamiltonian, the task of solving (\ref{eq:unitaryevolution}) reduces to computing the 
family of single-period time-evolution operators 
$$
U(t_0+T,t_0) = \mathcal{T} e^{-i\int_{t_0}^{t_0+T} dt' H(t')},
$$
where $0\leq t_0<T$. The various members of this family are unitarily equivalent: $U(t_0~+~T,t_0) = U^\dagger (0, t_0) U(T,0) U(0,t_0)$.

The eigenstates of $U(T)=U(T,0)$, 
\beq
U(T) |\phi_\alpha\rangle = e^{-i \epsilon_\alpha T} |\phi_\alpha\rangle,
\eeq
define special solutions of Eq.~(\ref{eq:unitaryevolution}): the Floquet eigenstates
\beq|\psi_\alpha (t)\rangle = U(t,0) |\phi_\alpha\rangle, \eeq 
which satisfy $|\psi_\alpha (t+T) \rangle = e^{-i \epsilon_\alpha T}|\psi_\alpha (t) \rangle$. Here the parameter $\epsilon_\alpha$ is called the quasienergy. The Floquet eigenstates explicitly exhibit the temporal periodicity of the Hamiltonian and form a basis for  general time evolution~\cite{Shirley65, Sambe1973}. The choice of quasienergy $\epsilon_\alpha$ is not unique as $\epsilon_\alpha \equiv \epsilon_\alpha + n_\alpha (2 \pi/T)$, where $n_\alpha$ can be any integer. This multivaluedness is related to the freedom in choosing the operator logarithm in $U(T)=e^{-i H_F T}$, to obtain what is called the Floquet Hamiltonian, $H_F$, (which is a Hermitian operator, but may not otherwise look much like a Hamiltonian.) Indeed, an alternative
formulation of the above is Floquet's theorem \cite{Floquet1883} which asserts that the evolution operator of a periodically driven system can be decomposed in the form 
\begin{equation}
  \label{eq:U_decomp}  U(t) = \Phi(t) e^{-i H_{F} t}, \quad \Phi(t) = \Phi(t + T),
\end{equation} 
where $\Phi(t)$ is periodic with the same period as the drive, $T$. The boundary condition $U(0) = \mathbb{1}$ implies that $\Phi(0) = \mathbb{1}$.

The Floquet effective Hamiltonian $H_{F}$ captures the ``stroboscopic'' dynamics of the system: given a state at time $nT$, where $n$ is an integer, the state at time $(n + 1)T$ can be obtained either by 1) evolving the system with the full time-dependent Hamiltonian $H(t)$ for $nT \le t < (n + 1) T$, or 2) evolving the system with the {\it time-independent} effective Hamiltonian $H_{F}$ for duration $T$. The reader is warned, though, not to assume that $H_{F}$ will always look like a standard Hamiltonian---it will not!

Before embarking on the technical discussions of the next few sections, we remark that care must be taken to distinguish between Floquet eigenstates, and the physical states realized in a given physical system.
The Floquet eigenstates form a complete basis for the single-particle Hilbert space; {\it any} evolution can therefore be expressed in this basis. Due to the inherently out-of-equilibrium nature of periodically driven systems, there is in general no guarantee that the state of a given system will have a simple description in terms of these eigenstates. Hence we will need to ask in each Floquet system: What initial states can be prepared? What late time states do they lead to? What signatures do they exhibit? How do we relate these signatures to properties of the Floquet eigenstates or of the drives as a whole? In some cases complete answers are known but in others these are still partially open questions.

%%%%%%%%%%%%%%%%%%%%%%%%%%%%%%%%%%%%%%%%%%%%%%%%%%%%

%%%%%%%%%%%%%%%%%%%%%%%%%%%%%%%%%%%%%%%%%%%%%%%%%%%%
\section{Topology of Single-particle Floquet Systems}
\label{sec:single_particle}

In this section we focus on the topological properties of periodically driven single-particle quantum systems.
In the presence of a crystal lattice, and with a uniform drive such that the system has discrete translation symmetry, crystal momentum remains a good quantum number.
The single-particle Floquet spectrum of a system with discrete translation symmetry is therefore organized into Floquet-Bloch bands analogous to those of electrons in crystalline solids. We note that since the single-particle properties that we discuss in this section are essentially features of a linear wave equation, many of the concepts can be extended to other physical settings such as electromagnetic waves~\cite{Rechtsman2013, Hu2015, Mukherjee2017, Maczewsky2017, Mukherjee2018} and mechanical systems~\cite{Peng2016} described by such equations. 

%%%%%%%%%%%%%%%%%%%%%%%%%%%%%%%%%%%%%%%%%%%%%%%%%
\subsection{Topology in Space {\it and} Time}

%%%%%%%%%%%%%%%%%%%%%%%%%%%%%%%%%%%%%%%%%%%%%%%%%%
\begin{figure*}[ht]
\includegraphics[width = 4.5 in]{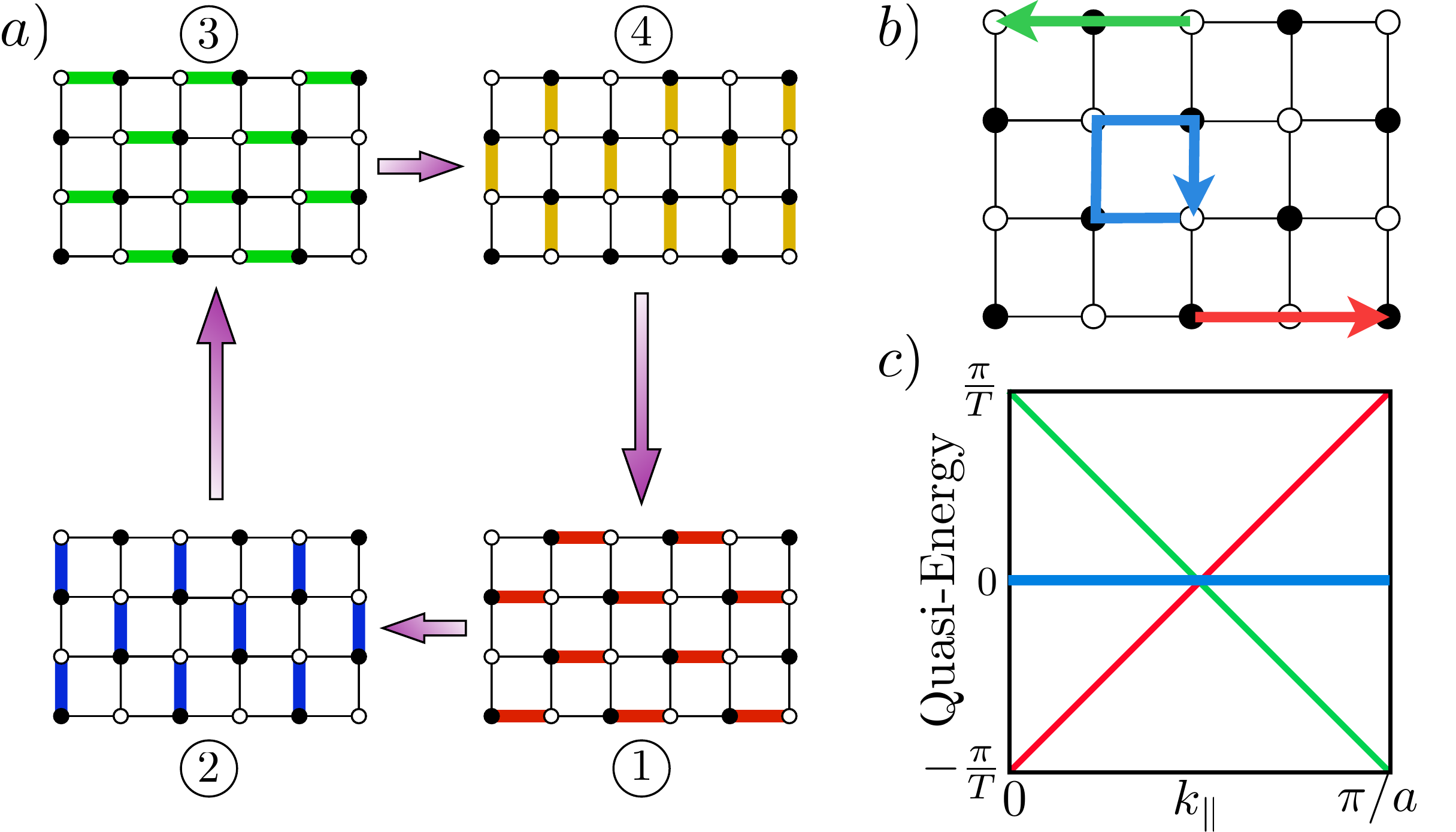}
\caption{Four-step tight-binding model illustrating the inadequacy of an effective Hamiltonian based approach to the topological characterization of single-particle Floquet systems [figure adapted from Phys.~Rev.~X {\bf 3}, 031005 (2013)].}
\label{fig:AFAI}
\end{figure*}
%%%%%%%%%%%%%%%%%%%%%%%%%%%%%%%%%%%%%%%%%%%%%%%%%%

After the discovery of topological insulators and the associated single-particle theory \cite{HasanTI_RMP}, it was natural to seek an understanding of the role of band topology in Floquet systems. Early approaches to the topological characterization of single-particle Floquet systems focused on classifying the Floquet-Bloch band structure associated with the effective Hamiltonian, $H_{F}$ [see, for example, the discussion in Ref.~\citenum{Oka2019}].
In this way, the vast body of results on the topological characteristics of Bloch bands in nondriven systems \cite{HasanTI_RMP, Xiao2010} could be directly imported to the Floquet setting. We will refer to such drives---wherein closely analogous physics is also present in undriven systems---as Type I. While Type I drives do not lead to qualitatively new physics, they can certainly lead to striking effects wherein $H_F$ can be ``engineered'' to exhibit different physics than the undriven $H$~\cite{Oka2009, Kitagawa2010, Lindner2011}. An example is the realization of 
nontrivial Chern bands in cold atomic experiments \cite{Aidelsburger2013, Jotzu2014, Flaschner2016}. 
There is by now a large body of literature on such Type-I Floquet topological systems, which we do not attempt to review in full here; this subject has been reviewed elsewhere, e.g., in Refs.~\citenum{Cayssol2013, Oka2019}. 
Here, our focus is primarily on what we will call Type~II Floquet drives, whose physical manifestations 
can {\it only} appear in the driven setting. (In the previous literature, Type II drives have been variously described as ``anomalous,'' ``inherently dynamical,'' or ``nontrivial.'')  
Somewhat later we will see that in the many-body localized setting these drives lead to entirely new phases of matter but for now we remain in the single-particle setting and describe the basics of this new physics.

One of the most striking consequences of nontrivial band topology is the existence of robust, topologically protected surface modes at interfaces between regions hosting bands with different topological indices.
Notable examples include the chiral edge modes of two-dimensional integer quantum Hall systems \cite{Halperin82} and the ``Majorana zero modes'' at the ends of one-dimensional topological superconductors \cite{Kitaev_2001}.
Formally, in each case the appearance of such modes is captured by a ``bulk-edge correspondence.''

Like their equilibrium counterparts, Floquet systems may also support topologically protected edge modes. 
Interestingly, the topological characteristics of the system's effective Hamiltonian $H_{F}$ do {\it not} uniquely determine its topological edge properties.
To illustrate why this is so, consider a two-dimensional periodically driven system, whose Floquet bands are characterized by a set of Chern numbers.
According to spectral-flow arguments \cite{Rudner2013,AFAI}, the Chern number of a given band is equal to the difference between the net chirality of edge modes traversing the gaps above and below the band (when the system is defined in a geometry with an edge connected to the vacuum).
In equilibrium, the spectrum is bounded from below and there cannot be any chiral edge states below the bottom of the lowest band. 
Therefore, knowing the Chern number of each band is sufficient to uniquely determine the net chirality of edge modes in every gap in the spectrum.
In a Floquet system, however, the spectrum is {\it periodic} in quasienergy: there is no notion of the `lowest' band, and therefore no zero from which to start the counting the chirality of edge modes.
Crucially, this implies that the Chern numbers of the Floquet bands do {\it not} provide sufficient information to predict the absolute chirality of the edge modes expected within each gap of the Floquet spectrum.  
In fact, as we show explicitly by example in the next subsection, this information is not contained in $H_{F}$ at all---to resolve the ambiguity, we must go beyond the stroboscopic picture and examine the continuous evolution over all times within the driving period.

\subsection{The RLBL Model \label{sec:RLBL_model}}

The inadequacy of an effective Hamiltonian-based topological characterization of Floquet systems can be simply illustrated through the exactly solvable RLBL model~\cite{Rudner2013}.
Consider a two-dimensional square lattice with periodically modulated nearest-neighbor hopping amplitudes as shown in Fig.~\ref{fig:AFAI}.
We identify two sublattices, $A$ and $B$, indicated by open and filled circles, respectively.
The hopping amplitudes are modulated through a piecewise constant, four-step driving protocol.
During each step $i = \{1, \ldots, 4\}$, of duration $T/4$ (such that $T$ is the driving period), a time-independent Hamiltonian $H_i$ is applied to the system.
We distinguish four different types of bonds, as indicated by the colors in Fig.~\ref{fig:AFAI}a.
In each step, as labeled by the numbers 1 to 4, the hopping amplitudes on the highlighted bonds are set to a value $J$, while all other hopping amplitudes are set to zero.
The value of $J$ is picked such that $J T/4 = \pi/2$, so that a particle initialized on a given site will be transferred to the neighboring site along the highlighted bond with unit probability over one step.

Consider a particle initialized on any site in the bulk of the system at the beginning of a driving period.
As shown in Fig.~\ref{fig:AFAI}b, over one driving cycle the particle hops around a plaquette, returning precisely to its initial state at the end of the driving cycle.
For a system with periodic boundary conditions, this perfect return after each cycle is realized for {\it any} initial state of the particle.
Thus the Floquet operator describing the stroboscopic evolution of the system is identity: $U(T) = \mathbb{1}$.
From the stroboscopic point of view, it appears that the system does not evolve at all; correspondingly, the Floquet Hamiltonian [see Eq.~(\ref{eq:U_decomp})] vanishes, $H_{F} = 0$.

Now consider a particle initialized on one of the $A$ sites (indicated by an open circle) on the upper edge of a system with open boundary conditions as depicted in Fig.~\ref{fig:AFAI}b.
Over one complete driving cycle, the particle hops two sites (one unit cell) to the left. 
Over subsequent cycles, the particle will continue to hop one unit cell to the left over each driving period.
A particle initialized on one of the sites indicated by a filled circle on the upper edge simply traverses a closed loop (similar to the behavior in the bulk).
Thus we see that the upper edge of the system hosts a left-moving chiral edge state with support on the $A$ sublattice.
Similar considerations show that the lower edge hosts a right-moving chiral edge state with support on the $B$ sublattice.

The Floquet spectrum for this system in a strip geometry is shown in Fig.~\ref{fig:AFAI}c.
The bulk states that return to themselves after each period comprise a doubly degenerate flat band at quasi-energy zero (indicated in blue).
The right- and left-moving chiral modes are indicated in red and green, respectively.
Due to the fact that the edge modes span a wide gap in the quasi-energy spectrum, their existence is robust against small perturbations that destroy the flat-band condition in the bulk (without closing the quasi-energy gap).
Thus, although we arrived at this situation by considering a fine-tuned exactly solvable model, we can conclude that there is a stable ``phase'' in which the system hosts topologically-protected chiral edge states, despite the fact that its effective Hamiltonian appears to be completely trivial.

The failure of the effective Hamiltonian to correctly capture the topological nature of the system is rooted in the fact that the stroboscopic picture ignores the {\it micromotion} \cite{Goldman2014}, i.e., the continuous evolution that takes place within each driving period. 
In the four-step drive example described above, $\Phi(t)$ stores the information about the chirality of the drive (i.e., whether particles circle around the plaquettes clockwise or counterclockwise), which in turn ultimately determines the chirality of the edge states on each edge.
The micromotion is captured by the unitary operator $\Phi(t)$ in Eq.~(\ref{eq:U_decomp}).
Due to the boundary conditions $\Phi(0) = \Phi(T) = \mathbb{1}$, $\Phi(t)$ is an example of a ``unitary loop:'' a unitary evolution operator that starts and ends at the identity.
Note that $\Phi(t)$ is $T$-periodic for {\it any} Floquet system, not just for the special model described above.

For a two-dimensional (2D) translation-invariant lattice system with periodic boundary conditions, $\Phi(t)$ is block-diagonal in crystal momentum, $\vec{k} = (k_x, k_y)$.
Labeling the block within the crystal momentum $\vec{k}$ sector $\Phi(k_x, k_y, t)$, we note that $\Phi(k_x, k_y, t)$ is a unitary operator that is periodic in all three of its arguments.
As pointed out in Ref.~\citenum{Rudner2013}, the micromotion operators of two-dimensional Floquet systems can thus be characterized by the topological winding number~\cite{Bott1978}:
\begin{equation}
    \label{eq:WindingNumber} \mathcal{W}[\Phi] = \frac{1}{8\pi^2}\oint \dd t \dd k_x \dd k_y\, {\rm Tr}\left(\Phi^\dagger \partial_t \Phi \left[\Phi^\dagger \partial_{k_x} \Phi, \Phi^\dagger \partial_{k_y} \Phi\right]\right).
\end{equation}
Indeed, a bulk-edge correspondence for 2D Floquet systems can be formulated in terms of the winding number index (\ref{eq:WindingNumber}), see Ref.~\citenum{Rudner2013} for details. We will return to the experimental observability of this single-particle physics below. Immediately we will describe generalizations of the physics discussed here that systematically take account of symmetry and dimensionality. 

%%%%%%%%%%%%%%%%%%%%%%%%%%%%%%%%%%%%%%%%%%%%%%%%%
\subsection{Floquet Topological Insulators: Symmetries and Classification}
%%%%%%%%%%%%%%%%%%%%%%%%%%%%%%%%%%%%%%%%%%%%%%%%%%
\begin{figure}[t]
\includegraphics[width = 3 in]{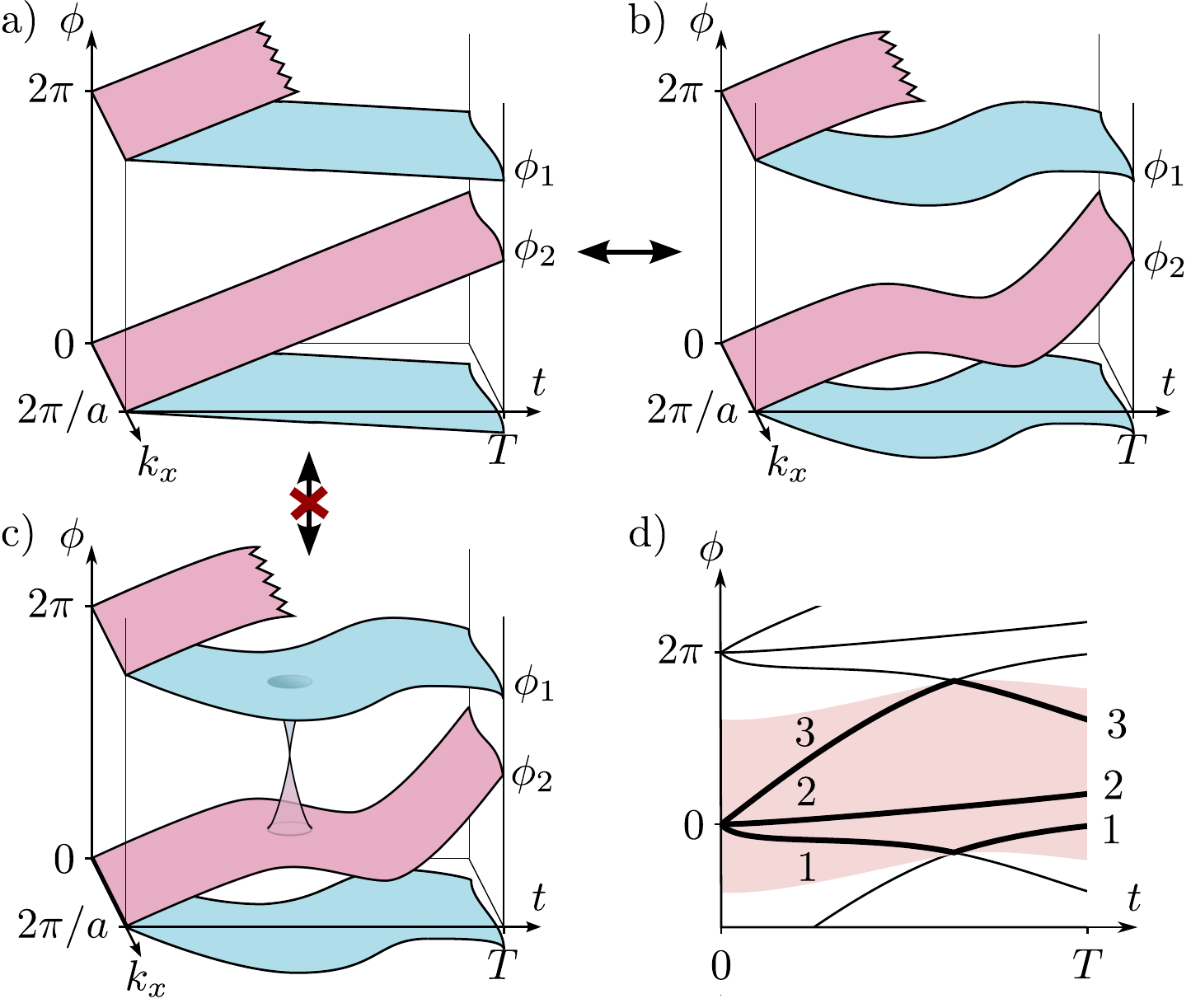}
\caption{Phase band picture of time evolution in single-particle Floquet systems [figure from New J.~Phys.~{\bf 17} 125014 (2015)].}
\label{fig:PhaseBands}
\end{figure}
%%%%%%%%%%%%%%%%%%%%%%%%%%%%%%%%%%%%%%%%%%%%%%%%%%

The results above demonstrate that both the effective Hamiltonian $H_{F}$ and the micromotion operator $\Phi(t)$ play crucial roles in determining the topological characteristics of single-particle Floquet systems.
The winding number in Eq.~(\ref{eq:WindingNumber}) distinguishes different ``micromotion phases.''
Importantly, the winding number must vanish for any nondriven system~\cite{Rudner2013}.
Therefore any evolution characterized by a nonzero value of $\mathcal{W}$ is topologically distinct from the evolution of any system with a static Hamiltonian.

The RLBL model is our first example of a Type II Floquet Topological Insulator (FTI). The designation ``topological insulator'' is inherited from the static case and includes systems that are better thought of as topological superconductors.
Since the generating Hamiltonian of the RLBL model did not require any symmetries (other than translational symmetry in the space and time dimensions), the system belongs to class~A of the Altland-Zirnbauer (AZ) symmetry classification \cite{Zirnbauer1996,Altland1997,Heinzner2005}.
In the static case, systems in this class may be associated with an integer topological invariant, corresponding to the total Chern number of the filled bands \cite{TKNN1982}. We see from Eq.~(\ref{eq:WindingNumber}), however, that Floquet systems in this class can be associated with an additional integer invariant corresponding to a winding number. 

In this way, the well-known `periodic table' of topological insulators and superconductors \cite{Kitaev2009}, which catalogues the topological phases of static  free-particle systems according to their symmetry class, must be extended in the periodically driven case. For static systems, the relevant discrete symmetries that define the symmetry classes are time reversal ($\mathcal{T}$), particle-hole conjugation ($\mathcal{C}$) and chiral (or sublattice) symmetry ($\mathcal{S}$). For a time-dependent system, these can be taken to act on the instantaneous Hamiltonian and time-evolution operator as
\begin{widetext}
\beq
\mathcal{T}H(\vec{k},t)\mathcal{T}^{-1}=H(-\vec{k},T-t)&~~~\to~~~&\mathcal{T}U(\vec{k},t)\mathcal{T}^{-1}=U(-\vec{k},T-t)U^\dagger(-\vec{k},T)\nonumber\\
\mathcal{C}H(\vec{k},t)\mathcal{C}^{-1}=-H(-\vec{k},t)&~~~\to~~~&\mathcal{C}U(\vec{k},t)\mathcal{C}^{-1}=U(-\vec{k},t)\label{eq:symmetry_defs}\\
\mathcal{S}H(\vec{k},t)\mathcal{S}^{-1}=-H(\vec{k},T-t)&~~~\to~~~&\mathcal{S}U(\vec{k},t)\mathcal{S}^{-1}=U(\vec{k},T-t)U^\dagger(\vec{k},T),\nonumber
\eeq
\end{widetext}
where $\mathcal{T}$ and $\mathcal{C}$ are antiunitary operators and $\mathcal{S}$ is unitary. We note that other generalizations of the symmetry operator actions to the Floquet case can be used instead, but that the conclusions remain unchanged. The presence or absence of these symmetries, and whether the antiunitary symmetries square to $\pm1$, define the 10 AZ symmetry classes (see Table.~\ref{tab:FTI_periodic_table} for precise definitions).

In a series of pioneering early works, models for Type II FTI phases were obtained for 1D chains with emergent Majorana fermions \cite{Jiang2011,Liu2013,Thakurathi2013,Kundu2013,Reynoso2013,Thakurathi2017}, in 2D systems with time-reversal symmetry \cite{Carpentier2015}, in 1D systems with chiral symmetry \cite{Asboth2014}, and in complex symmetry classes in all dimensions \cite{Fruchart2016}, in addition to the RLBL model introduced above \cite{Rudner2013}. Several other works also considered topological phases of driven systems in the context of quantum walks [see, e.g., \cite{Kitagawa2010a, Kitagawa2012, Asboth2012,Tarasinski2014,GNVW}] and adiabatic cycles \cite{Zhang2014}, which may be interpreted as FTI phases. 

A universal classification of FTI phases, analogous to the static periodic table, requires a detailed study of the different sets of micromotion operators $\Phi(t)$ that are possible within each symmetry class. This question has been addressed from two complementary perspectives, which we now discuss.

%%%%%%%%%%%%%%%%%%%%%%%%%%%%%%%%%%%%%%%%%%%%%%%%%
\subsection{Phase Bands}

Insight into the nature of quantum evolution in Floquet systems, and how it may differ from evolution in nondriven systems, 
can be gained by examining the ``phase bands'' of the system \cite{Nathan2015}.
In a system with $N$ bands, the phase bands $\phi_n(\vec{k},t)$ are defined via the spectral representation of the Fourier space evolution operator $U(\vec{k},t)$:
\begin{equation}
    \label{eq:PhaseBands} U(\vec{k}, t) = \sum_{n=1}^N P_n(\vec{k}, t) e^{-i \phi_n(\vec{k}, t)}, 
\end{equation}
where $P_n(\vec{k}, t)$ is the projector onto the $n$-th eigenstate of $U(\vec{k},t)$ at time $t$.
As shown in Fig.~\ref{fig:PhaseBands}, the phase bands provide a visualization of how the evolution builds up from the identity at $t = 0$ to the Floquet operator at $t = T$: $U(\vec{k},T) = \sum_{n=1}^N P_n(\vec{k}, T) e^{-i \epsilon_n(\vec{k})T}$, where $\epsilon_n(\vec{k}) T = \phi_n(\vec{k}, T)$.

Due to the initial condition $U(\vec{k}, 0) = \mathbb{1}$, at $t = 0$ the phases $\phi_n(\vec{k}, 0)$ must all be integer multiples of $2 \pi$.
For a system with a time-independent Bloch Hamiltonian $H_{\rm static}(\vec{k})$, with spectrum $H_{\rm static}(\vec{k})\ket{n, \vec{k}} = E_n(\vec{k})\ket{n, \vec{k}}$, the projectors $\{P_n(\vec{k}) = \ket{n, \vec{k}}\bra{n, \vec{k}}\}$ are time-independent and the phases wind linearly in time: $\phi_n(\vec{k}, t) = E_n(\vec{k}) t$.
The phase bands thus form linear sheets as shown in Fig.~\ref{fig:PhaseBands}a.
For a system with a time-dependent Hamiltonian, the phase bands may develop more interesting structure.
In the example shown in Fig.~\ref{fig:PhaseBands}b, the phase band sheets may be continuously deformed/straightened out; thus the evolution is smoothly connected to the evolution of a nondriven system.
Importantly, as shown in Fig.~\ref{fig:PhaseBands}c, phase bands from neighboring ``phase Brillouin zones'' may intersect at topologically protected singularity points.
For a (2 + 1)-dimensional system, these singular points are similar to the Weyl points in 3D band structures and cannot be removed by any small perturbation of the underlying Hamiltonian of the system.

There is no smooth deformation of the phase bands in Fig.~\ref{fig:PhaseBands}c that can make the evolution look like that of a nondriven system (Fig.~\ref{fig:PhaseBands}a).
Similar to presence of a nonvanishing winding number $\mathcal{W} \neq 0$ in Eq.~(\ref{eq:WindingNumber}), the presence of topological singularities in the phase band structure signals that an evolution is  qualitatively distinct from that of any nondriven system. 
Indeed, by systematically studying the general form of such topological singularities, it is shown in Ref.~\citenum{Nathan2015} that 2D Floquet systems belonging to class~A are classified by $N$ integers (or elements from the group $\mathbb{Z}^{\times N}$), where $N$ is the number of bands in the Floquet Hamiltonian $H_{F}$. This is to be contrasted with the $(N-1)$ independent Chern numbers that describe a static Hamiltonian with $N$ bands. Ref.~\citenum{Nathan2015} demonstrates that, in this way, the bulk-edge correspondence for such systems can equivalently be formulated in terms of the winding number $\mathcal{W}$ or directly in terms of phase band topological singularities and the Chern numbers of $H_{F}$.

The phase band picture provides a natural formalism for studying dynamical topological evolutions from any of the 10 AZ symmetry classes. In general, the expression for the unitary operator in terms of phase bands [Eq.~(\ref{eq:PhaseBands})] will satisfy additional symmetry constraints according to the symmetry operators defined in Eq.~(\ref{eq:symmetry_defs}). However, the role of topological singularities formed at the phase Brillouin zone boundaries remains the same: By studying the structure of these singularities in the presence of symmetry, a classification of the corresponding Floquet phases can be obtained. Notably, Ref.~\citenum{Nathan2015} applies this approach to 1D Floquet systems with particle-hole symmetry (in class~D), which are described by two parity indices, while 2D and 3D systems with time-reversal symmetry (in class~AII) are shown to be described by $N$ $\zbb_2$ indices, where $N$ is again the number of bands in $H_{F}$. In each of these cases, this adds one additional topological index to the classification of the corresponding static phases. The method of classifying singularities using the phase band paradigm may be generally extended to any dimension and symmetry class.

%%%%%%%%%%%%%%%%%%%%%%%%%%%%%%%%%%%%%%%%%%%%%%%%%
\subsection{Unitary Loops and the Periodic Table for Floquet Topological Insulators\label{sec:loops_periodic_table}}
A complementary approach to the classification of FTI phases \cite{Roy2017b} uses methods from K-theory \cite{Atiyah1967,Karoubi1978}, a branch of mathematics that was used previously to obtain the periodic table of \emph{static} topological insulators and superconductors (TIs) \cite{Kitaev2009}. The static periodic table gives the number of equivalence classes of gapped Hamiltonians that are related to each other homotopically, for each dimension and symmetry class. Specifically, two gapped Hamiltonians are connected by (stable) homotopy if one can be continuously deformed into the other without breaking any protecting symmetries, without closing the spectral gap (assumed to be at zero energy), and with the allowed addition of any number of trivial bands. The result is that for each dimension and symmetry class, the set of equivalence classes is in one-to-one correspondence with one of the Abelian groups $\{\zbb, \zbb_2, 0\}$. Topologically nontrivial phases then correspond to equivalence classes with a nonzero (integer or $\zbb_2$-valued) topological invariant.

In seeking a similar, homotopic classification of Floquet phases, we are faced with the problem that every unitary operator is connected to every other. Specifically, if we think of a unitary evolution $U(t)$ as tracing out a path in the space of all unitary operators, this can always be `wound back' to the identity, which is a manifestly trivial unitary operator. To add some distinction between different unitary evolutions, we will demand that a unitary evolution \emph{for a system without a boundary} has gaps in the quasienergy spectrum at its endpoint. Without loss of generality \cite{Roy2017b}, we will specifically assume that there are gaps at $\epsilon=0$ and at $\epsilon=\pi/T$. Note that this implies that a Floquet Hamiltonian $H_{F}$ may be defined that has all its energies satisfying $0<|E|<\pi/T$. We make this restriction for the system without a boundary because a system with an edge may host boundary modes that lie in these quasienergy gaps.\footnote{The gap requirement is natural for systems with translational symmetry (which we consider here), but can be generalized to a mobility gap in systems with disorder (see discussions in Refs.~\citenum{Roy2017b,Liu2018,Graf2018}).} Two unitary evolutions are then homotopically connected if one can be deformed into the other without closing the quasienergy gaps and without breaking any protecting symmetries, if present.

\begin{figure}
\center
\includegraphics[scale=0.3]{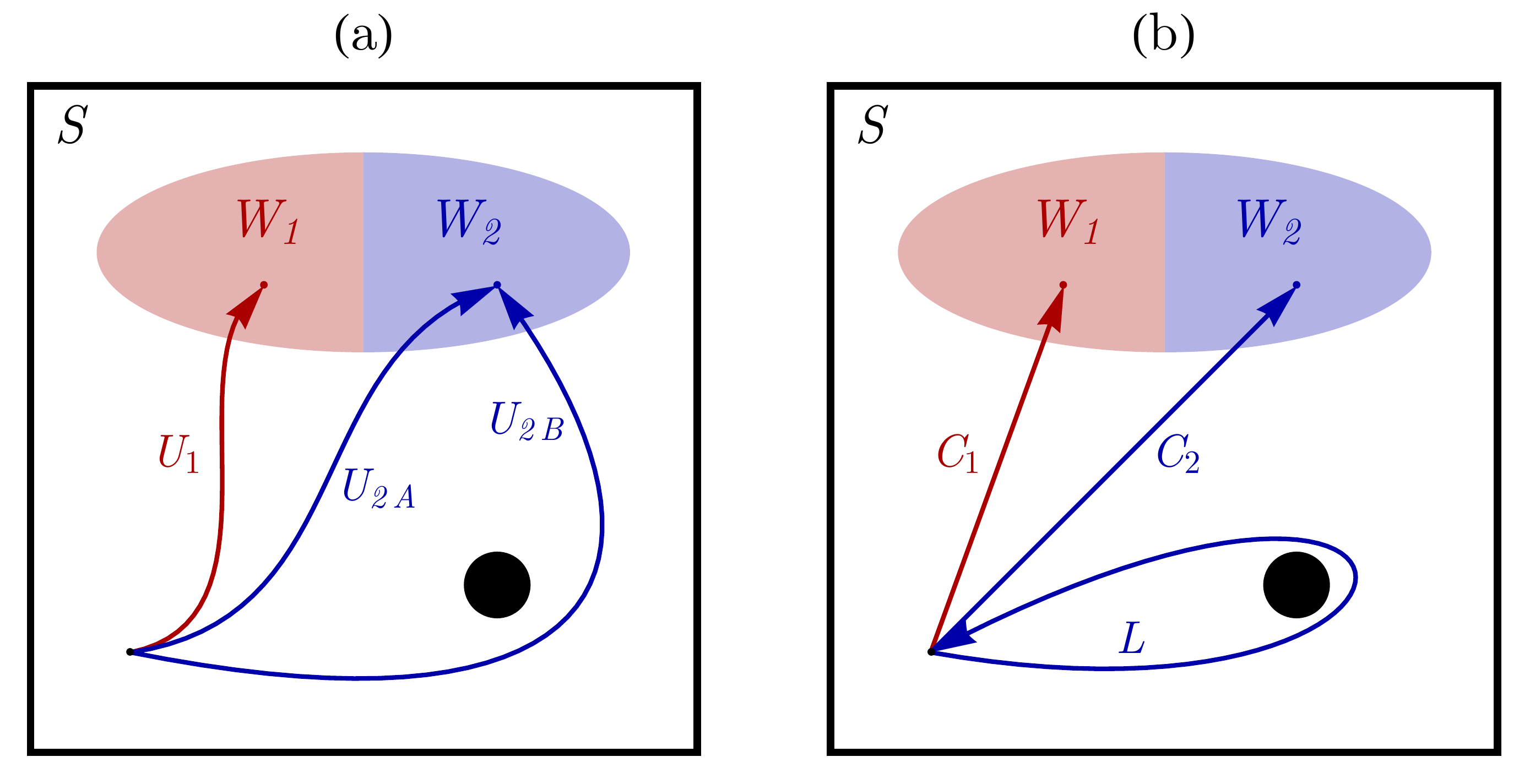}
\caption{The space of (single-particle) unitary operators $S$, containing two distinct gapped endpoint regions $W_1$ and $W_2$. An obstruction is indicated as a large black disk. (a) Path $U_1$ ends in a different endpoint region to paths $U_{2A}$ and $U_{2B}$, and so is homotopically distinct. Paths $U_{2B}$ and $U_{2A}$ both end in $W_2$ but are homotopically distinct because one cannot be continuously deformed into the other without crossing the obstruction. (b) Unitary $U_{1}$ and $U_{2A}$ can be homotopically deformed into the constant evolutions $C_1$ and $C_2$, respectively. However, unitary $U_{2B}$ can only be homotopically deformed into the composition $L*C_2$. This figure can also be interpreted in the many-body setting, where $S$ is now the space of many-body unitary operators and $C_1$ and $C_2$ are canonical paths.\label{fig:unitary_phase_space}}
\end{figure}

This notion of homotopy leads to two distinct types of equivalence, illustrated schematically in Fig.~\ref{fig:unitary_phase_space}(a). In the first case, two unitary evolutions may be distinct because they correspond to distinct end point regions (or equivalently, distinct classes of Floquet Hamiltonians). For example, in Fig.~\ref{fig:unitary_phase_space}(a), regions $W_1$ and $W_2$ might correspond to two sets of Floquet Hamiltonians with different Chern numbers. $U_1$ cannot be deformed into $U_{2A}$ without crossing the boundary between these regions, which would close the quasienergy gap at $\epsilon=0$. This partition of endpoint regions captures the different equivalence classes of gapped Floquet Hamiltonians. However, since $H_{F}$ is a static quantity, it cannot describe the Type II FTI phases that can only arise in driven systems. 

The second type of equivalence describes different kinds of micromotion, as introduced in Sec.~\ref{sec:basic_formalism}, which may be different even for two unitary evolutions with the same Floquet Hamiltonian. In Fig.~\ref{fig:unitary_phase_space}(a), this is demonstrated by unitary evolutions $U_{2A}$ and $U_{2B}$, which correspond to the same $H_{F}$ but which cannot be continuously deformed into one another because of the obstruction in the space of unitaries (represented by the black disk). Crossing this obstruction would close the quasienergy gap at $\epsilon=\pi/T$.

To classify this latter type of obstruction, it is useful to deform a general (gapped) unitary evolution into a sequence of two components: a unitary loop $L$, which is an evolution satisfying $L(0)=L(T)=\id$, followed by an evolution $C$ with constant Hamiltonian $H_{F}$. We write this deformation as $U\to L*C$, where $*$ indicates a temporally sequential composition of unitary evolutions \cite{Roy2017b} (see schematic illustration in Fig.~\ref{fig:unitary_phase_space}(b)).\footnote{Note that this deformation appears similar to the decomposition involving the micromotion operator described above, but in this case the loop evolution $L$ occurs sequentially in time \emph{before} the constant evolution $C$; thus here we truly deform the evolution, while the decomposition in Eq.~(\ref{eq:U_decomp}) is simply an exact rewriting of the  original evolution.} In Ref.~\citenum{Roy2017b}, it is proved that any gapped unitary evolution can be uniquely decomposed in this manner (preserving any symmetries if required).

A classification of Type~II FTI phases amounts to a classification of unitary loops $L(t)$. Although loop evolutions may seem somewhat fine-tuned, they should be interpreted as the loop component of a more general (and physically realistic) evolution, following the deformation $U\to L*C$. To classify equivalence classes of loops, we use the mapping
\beq
L(\vec{k},t)&\to&H_L=\left(\begin{array}{cc}
0 & L(\vec{k},t)\\
L^\dagger(\vec{k},t) &0
\end{array}\right),
\eeq
which takes the unitary loop to a gapped Hermitian operator. This Hermitian operator can be interpreted as a static `Hamiltonian' in $(d+1)$ dimensions, generally with additional symmetries due to the doubled system size \cite{Roy2017b}. By applying a sequence of K-theory identities \cite{Atiyah1967,Karoubi1978}, the K-group that classifies the equivalence classes of such operators can be obtained. The end result is that unitary loop evolutions have the same Abelian classifying group $\{\zbb,\zbb_2,0\}$ as static TIs from the same symmetry class (with the original number of spatial dimensions) \cite{Roy2017b}.

\begin{table*}[t!]
\be
\arraycolsep=6pt\renewcommand\arraystretch{1.1}\begin{array}{c|ccc|cccccccc}
\hline\hline
{\rm AZ~Class} & \mathcal{T} & \mathcal{C} & \mathcal{S} & d=0 & 1 & 2 & 3 & 4 & 5 & 6 & 7 \\
\hline\hline
{\rm A}& 0 & 0 & 0 & \mathbb{Z}^{\times N}& \emptyset & \mathbb{Z}^{\times N}& \emptyset & \mathbb{Z}^{\times N}& \emptyset & \mathbb{Z}^{\times N}& \emptyset \\
{\rm AIII}  & 0 & 0 & 1 &  \emptyset & \mathbb{Z}^{\times 2}& \emptyset & \mathbb{Z}^{\times 2}& \emptyset & \mathbb{Z}^{\times 2}& \emptyset & \mathbb{Z}^{\times 2}\\
\hline
 {\rm AI} & + & 0 & 0 & \mathbb{Z}^{\times N}& \emptyset & \emptyset & \emptyset & \mathbb{Z}^{\times N}& \emptyset & \mathbb{Z}_2^{\times N} & \mathbb{Z}_2^{\times N} \\
{\rm BDI} & + & + & 1 &  \mathbb{Z}_2^{\times 2} & \mathbb{Z}^{\times 2}& \emptyset & \emptyset & \emptyset & \mathbb{Z}^{\times 2}& \emptyset & \mathbb{Z}_2^{\times 2}\\
{\rm D} & 0 & + & 0 &  \mathbb{Z}_2^{\times 2} & \mathbb{Z}_2^{\times 2} & \mathbb{Z}^{\times 2}& \emptyset & \emptyset & \emptyset & \mathbb{Z}^{\times 2}& \emptyset \\
 {\rm DIII} & - & + & 1 & \emptyset & \mathbb{Z}_2^{\times 2} & \mathbb{Z}_2^{\times 2} & \mathbb{Z}^{\times 2}& \emptyset & \emptyset & \emptyset & \mathbb{Z}^{\times 2}\\
{\rm AII} & - & 0 & 0 & \mathbb{Z}^{\times N}& \emptyset & \mathbb{Z}_2^{\times N} & \mathbb{Z}_2^{\times N} & \mathbb{Z}^{\times N}& \emptyset & \emptyset & \emptyset \\
{\rm CII} & - & - & 1 & \emptyset & \mathbb{Z}^{\times 2}& \emptyset & \mathbb{Z}_2^{\times 2} & \mathbb{Z}_2^{\times 2} & \mathbb{Z}^{\times 2}& \emptyset & \emptyset \\
 {\rm C} & 0 & - & 0 & \emptyset & \emptyset & \mathbb{Z}^{\times 2}& \emptyset & \mathbb{Z}_2^{\times 2} & \mathbb{Z}_2^{\times 2} & \mathbb{Z}^{\times 2}& \emptyset \\
{\rm CI} & + & - & 1 & \emptyset & \emptyset & \emptyset & \mathbb{Z}^{\times 2}& \emptyset & \mathbb{Z}_2^{\times 2} & \mathbb{Z}_2^{\times 2} & \mathbb{Z}^{\times 2}\\
\hline\hline
\end{array}.
\ee
\caption{Periodic table for gapped FTI phases, arranged by symmetry class. The left four columns give the definition of each AZ class in terms of the presence/absence of symmetry operators and their square. The right eight columns give the topological classification for spatial dimension $d$, where $N$ is the number of bands in the Floquet Hamiltonian. The table repeats for $d\geq 8$ (Bott periodicity). \label{tab:FTI_periodic_table}}
\end{table*}

Since a general unitary evolution can be deformed into a loop evolution followed by an evolution with a (constant) Floquet Hamiltonian, the complete classification is a combination of each of these pieces. In symmetry classes without particle-hole or chiral symmetry, a system whose Floquet Hamiltonian has $N$ bands is classified by $N$ factors of the corresponding static classification: $N-1$ of these correspond to properties of $H_{F}$ (e.g. Chern numbers), while the final factor corresponds to the loop component. In systems with particle-hole or chiral symmetry, the classification is given by two factors of the corresponding static classification: in these cases, only the quasienergy gaps at $\epsilon=0$ and $\epsilon=\pi/T$ are physically meaningful, and so these are the only gaps that permit edge modes \cite{Roy2017b}. These results are summarised in Table~\ref{tab:FTI_periodic_table}. This table was applied to topological defects, and a complete set of topological invariants for each entry obtained, in Ref.~\citenum{Yao2017}

%%%%%%%%%%%%%%%%%%%%%%%%%%%%%%%%%%%%%%%%%%%%%%%%%%%%
\section{Many-body systems: Generalities}

Thus far we have discussed the behavior of strictly single-particle systems;  now we must face up to the reality that the particles involved must interact, at least weakly. In this section we will review two sets of necessary general ideas---the first have to do with the challenge of heating and the second have to do with the techniques needed to get an understanding of the topology of Floquet drives in many body systems.

\subsection{Heating, Localization and Eigenstate Order\label{sec:heating_etc}}

Naively, a closed, interacting Floquet system should heat to infinite temperature. One way to see this is to note that with energy conservation gone, the entropy maximizing state is the infinite temperature state \cite{Lazarides2014,DAlessio2014,Ponte2015b}, with all local operator expectation values  time independent at long times irrespective of the starting state. A second is to recall that linear response theory generically predicts absorption at any non-zero frequency short of infinite temperature; in an interacting system this energy absorbed should get redistributed over all degrees of freedom. If this were the full story the work on non-interacting drives we reviewed above would be of limited experimental interest, as any weak interaction would lead to all the interesting topology being buried under the noise of heating. Fortunately, this trivial outcome is {\it not} the only one possible. There are three known possibilities for evading it: many-body localization (MBL), prethermalization regimes, and cooling.

Of these, Floquet many-body localization provides the most complete realization of the idea of Floquet phases protected from heating. For these localized systems, disorder in the couplings leads to the emergence of $O(N)$ 
spatially localized, mutually commuting, `l-bit' operators $\tau_i^z$ (which  depend on details of the drive) such
that 
%\begin{equation}
$
    [U(T),\tau^z_i]=0.
$
%\end{equation}
Intuitively, the system breaks up into local Rabi oscillators which do not transfer energy to each other
but which do influence each other's phase; the individual oscillators simply absorb and transfer energy back to the driving system over each period. This behavior is stable at high frequencies (small $T$) but gives way to heating at low frequencies (large $T$). A combination of computational studies, along with more detailed qualitative and analytic arguments,
\cite{Ponte2015,Ponte2015b,Lazarides2015,Abanin2016} as well as very recent experimental work \cite{Bordia2017}, support the existence of the Floquet-MBL regime. 
We note that Floquet-MBL systems avoid heating generically---weak perturbations of Floquet-MBL
drives that leave the period unchanged are also Floquet-MBL.  

The lack of heating shows up as a breakdown of the Floquet analog of ETH---which states that individual Floquet eigenstates all yield trivial infinite temperature correlations. This breakdown now allows a sharp definition of ordering via the notion of eigenstate order \cite{Huse2013,Pekker2014} wherein we examine individual Floquet eigenstates for signatures of order. For example we can consider two-point correlation functions in a given Floquet eigenstate,
%\begin{equation}
    $\langle{\phi_\alpha|} O({\bf x_1},t_1) O({\bf x_2},t_2) | \phi_\alpha \rangle $
%\end{equation}
where the Heisenberg picture operators are evolved by means of the exact unitary time-evolution operators $U(t_{1,2},0)$ from a chosen reference time we set WLOG to zero.
Finally, Floquet-MBL systems, like Hamiltonian MBL systems, exhibit a degree of initial state independence at late times \cite{vonKeyserlingk2016b}. A general initial state 
$
|\chi(0) \rangle = \sum_\alpha c_\alpha |\psi_\alpha(0) \rangle
$
gives rise to the stroboscopic expectation values
\begin{widetext}
\begin{eqnarray}
\langle \chi(t+nT)| O |\chi(t+nT) \rangle =  
\sum\limits_\alpha \sum\limits_\beta
c_\alpha^* c_\beta e^{-i (\epsilon_\alpha -\epsilon_\beta)nT} 
\langle \psi_\beta (t)| O|\psi_\alpha(t) \rangle  \ ,
\end{eqnarray}
which at late times $n\gg1$ reduce to their values in the quasi-diagonal ensemble:
\begin{eqnarray}
\langle \chi(t+nT)| O |\chi(t+nT) \rangle \sim 
 \sum\limits_\alpha \sum\limits_{\beta(\alpha)} e^{{inT(\epsilon_\alpha-\epsilon_{\beta(\alpha)}})} c_\alpha^\ast c_{\beta(\alpha)}
 \langle \psi_\alpha(t)|O|\psi_{\beta(\alpha)}(t) \rangle  \ , 
\end{eqnarray}
\end{widetext}
where $\beta(\alpha)$ are members of special, ordering related multiplets that contain $\alpha$ which we will
discuss below. Thus at late times, roughly half the parameters present in the specification of the initial state (the phases) no longer influence local measurements. This allows eigenstate order to be detected dynamically---we simply start with a small class of initial easily prepared states and evolve them to long times, which is just as well as the preparation of eigenstates is not a feasible task for a many-body system. We refer the reader to Ref.~\citenum{Moessner2017} for a more expanded introduction to Floquet-MBL systems where the possibility of fine tuned integrable drives is also discussed.

The second counter to the heating problem is  a ``pre-thermal'' regime where heating simply takes a parametrically long time of $O(e^{J/T})$ in some characteristic local energy scale $J$. The intuition behind this is that it takes a very high order ``multi-photon'' process for absorption \cite{BukovReview,Abanin2015,Eckardt2015}. This regime has been rigorously shown to exist \cite{Abanin2017a,Kuwahara2016} for sufficiently high driving frequencies. It was shown subsequently that the prethermal regime is capable of supporting symmetry-breaking and symmetry-protected Type II Floquet phases \cite{Else2017} in addition to Type I phases where the original arguments sufficed. It is important to emphasize though that in many such cases only a vanishing fraction of the Floquet eigenstates exhibit ordering and so it is in general challenging to prepare initial states capable of detecting the order.

Finally we note that this review is focused, by design, on an idealized setting of an isolated many-body system, which is unaffected by coupling to its environment. 
In an experimental setting, such coupling will inevitably be present and may significantly alter the long-time trajectory of the system. While cold atom systems may be approximately described as being decoupled from their environments for significant lengths of time, the open system dynamics of Floquet systems is particularly relevant for experiments in solid state systems.
In particular, in the presence of a bath, a driven many-body system will generically come to a steady state at long times that need not be the completely trivial infinite temperature like state described above for thermalizing, closed, driven systems. The question for our current purposes is whether it can still reflect the physics discussed in this review. 
 
Determining the nature of the steady state of a Floquet system in the presence of dissipative coupling to a bath is a complex problem: unlike in equilibrium, where the steady state of a system is determined by just a handful of macroscopic parameters (temperature, chemical potential, etc.), out of equilibrium there is no detailed balance and the nature of the steady state becomes sensitive to microscopic details of the bath and the system-bath coupling~\cite{Kohn2001,Hone2009}.
Nonetheless, under appropriate conditions, it has been shown that Floquet topological insulator-like states may be stabilized by the action of a phonon bath on the driven system~\cite{Dehghani2014, Dehghani2015, Seetharam2015, Iadecola2015, Seetharam2018}.
The more general fate of intrinsically dynamical and strongly correlated Floquet phases in the presence of coupling to a bath remains an interesting and important direction for future research.

%%%%%%%%%%%%%%%%%%%%%%%%%%%%%%%%%%%%%%%%%%%%%%%%%%%%

\subsection{Many-body Unitary Loops and Edge Unitaries\label{sec:edge_unitaries}\label{sec:many_body_loops}}

As in the single-particle case, we can think of a many-body unitary evolution $U(t)$ (with periodic boundary conditions) as tracing out a path in the space of unitary operators (possibly with some symmetry requirement), as in Fig.~\ref{fig:unitary_phase_space}. Such a path will be continuous (or can be brought into a continuous form), but will not necessarily be smooth. 
To add some distinction between different unitary evolutions, we again place restrictions on their allowed endpoint regions, $W_j$. There are many meaningful choices for $W_j$ in the many-body case \cite{Roy2017}, but in this review we consider the physically relevant case where $W_j$ describe Floquet Hamiltonians that are many-body localized. In this case, two unitary evolutions are equivalent if they can be continuously deformed into one another without the Floquet Hamiltonian undergoing a delocalization transition. We note that this choice means that any symmetries, if present, must be compatible with many-body localization \cite{Potter2015,Potter2016b}. Different choices of endpoint region which lift this restriction are discussed in Ref.~\citenum{Roy2017}.

As before, an arbitrary many-body unitary evolution can be homotopically deformed into a loop evolution followed by an evolution $C$, which we now refer to a canonical path. For example, a canonical path might be a constant evolution with the Floquet Hamiltonian (although see Ref.~\citenum{Roy2017} for a discussion of the some of the subtleties involved in this definition). We write this deformation as $U\to L*C$, which is again represented schematically in Fig.~\ref{fig:unitary_phase_space}, now reinterpreted in a many-body context.

This decomposition splits the classification of many-body unitary evolutions into two components. The first is a classification of different endpoint regions $W_j$, which correspond to different sets of (MBL) Floquet Hamiltonians. For example, in the presence of symmetry there may be endpoint regions corresponding to different types of symmetry-protected topological (SPT) order, separated by delocalization transitions. This component is equivalent to a classification of static MBL systems, which have been well-studied elsewhere \cite{Huse2013,Bauer2013,Chandran2014,Pekker2014,Slagle2015,Potter2015,Bahri2015,Parameswaran2018}. 

The second, inherently dynamical component of the classification arises from obstructions in the space $S$, which again correspond to a classification of unitary loop evolutions. For this reason, we will devote much of the rest of this section to studying and classifying many-body unitary loop drives. Although they may seem somewhat fine-tuned, they should be interpreted as the loop component of a more general (and physically realistic) evolution which, for example, may be stabilized by including strong disorder.  This deconstruction allows us to separate questions about the stability of a drive from questions about its topology. 

Complementary to the description in terms of the bulk, a nontrivial Floquet evolution can be also identified by its edge action, i.e., the behaviour of the drive at the boundary of a system with one or more edges.
In many cases, it is useful to study this edge behaviour directly by defining an `edge unitary operator' $\hat{Y}$ that acts only in the vicinity of the boundary. For a loop evolution, the edge unitary may be extracted straightforwardly by evolving with the system Hamiltonian in the presence of open boundary conditions (OBC). Explicitly, the evolution decomposes at its endpoint into $U_{\rm OBC}(T)=I_{\rm bulk}\otimes \hat{Y}$, where $\hat{Y}$ is the component of $U_{\rm OBC}(T)$ that differs from the identity. As shown in Ref.~\citenum{Harper2017}, $\hat{Y}$ is limited in size by the Lieb-Robinson velocity \cite{lieb1972} of the generating Hamiltonian, and can be isolated straightforwardly from the identity contribution in the bulk. For evolutions which are not loops, the edge unitary $\hat{Y}$ is more difficult to extract, but can be obtained following the method in Ref.~\citenum{Po2016} if the endpoint exhibits MBL.

\section{Many-body Systems: Symmetry Breaking and Topology}
\label{sec:many_body_systems}
%%%%%%%%%%%%%%%%%%%%%%%%%%%%%%%%%%%%%%%%%%%%%%%%%%%%

\begin{figure}[h]
\includegraphics[width=0.70\columnwidth]{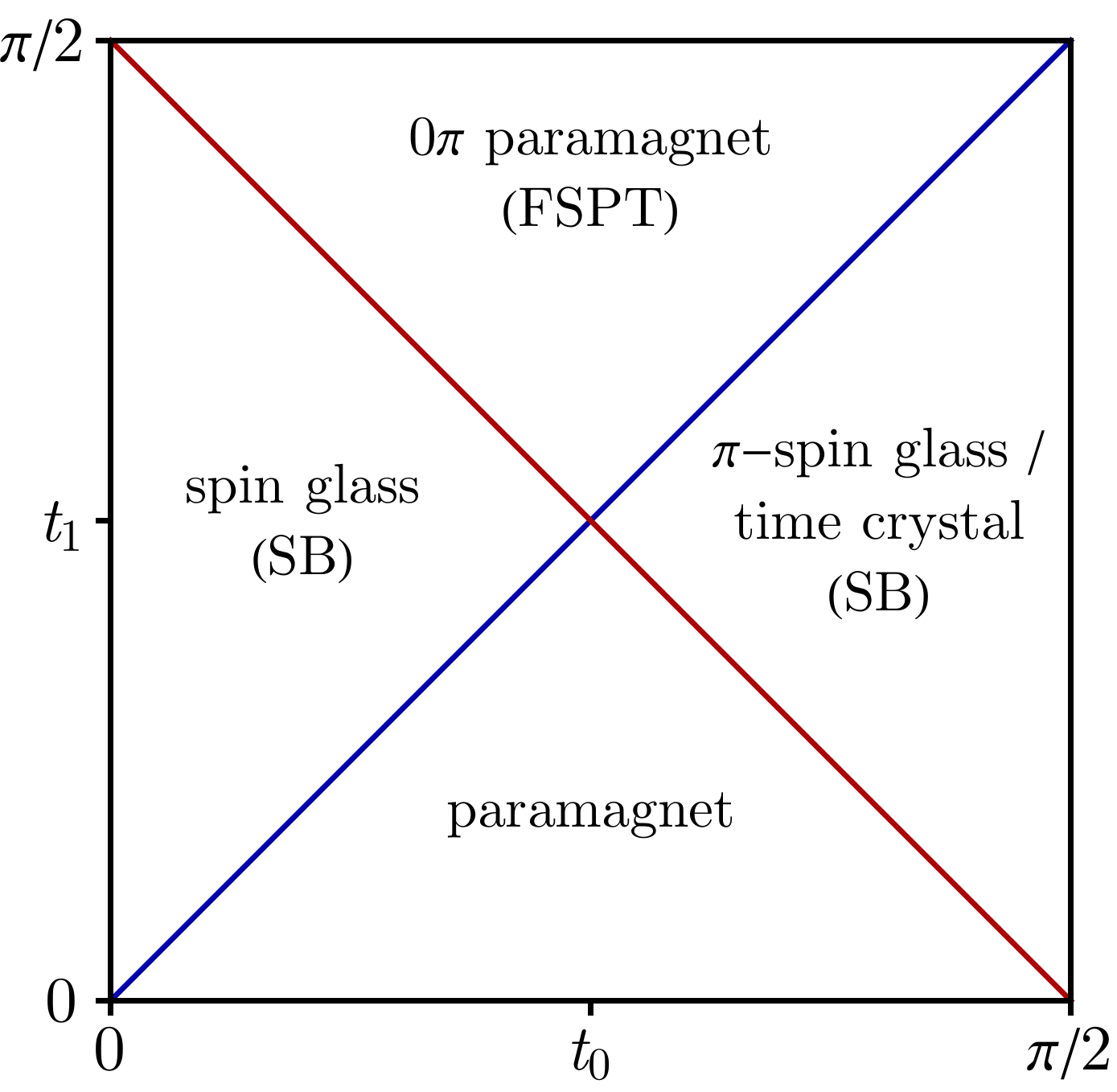}
\caption{This shows the phase diagram for the binary drive of Eq.~(\ref{eq:classDUf}). The red and blue lines separate distinct Floquet phases.} 
\label{classDPhasediagram}
\end{figure}

As we begin our concrete discussion of many-body Floquet systems we introduce a second simple model---literally and figuratively the Ising model of the phenomena we are exploring. It will exhibit the physics 
of topology {\it and} of symmetry breaking. 

This model \cite{Khemani2016,vonKeyserlingk2016a} consists of a binary drive:
\begin{equation}
\label{eq:classDUf}
U(t)=
\begin{cases}
e^{-iH_{0}t} & 0\leq t<t_{0}\\
e^{-iH_{1}(t-t_{0})}e^{-iH_{1}t_{0}} & t_{0}\leq t<t_{0}+t_{1},
\end{cases}    
\end{equation}
where
\beq
H_{0}=  -\sum_s  h_{s}X_s, &~~~~~~& H_{1}=  -\sum_s J_{s}Z_s Z_{s+1},
\eeq
and $X,Z$ are Pauli-matrix operators.
The parameters $h_{s}$, $J_{s}$ are made random to obtain localization, but for the 
purposes of obtaining simple phase boundaries will be taken to be (almost) spatially uniform. Both Hamiltonians commute with a $\mathbb{Z}_2$ global Ising symmetry operator $P=\prod_s X_s$.  This family of drives leads to the phase diagram shown in Fig.~\ref{classDPhasediagram}. Of the phases shown there, two are Type I Floquet continuations of static phases: the undriven paramagnet with Ising symmetric eigenstates, and the undriven spin glass with Ising symmetry-breaking eigenstates. The remaining two Type II phases are new. The first is a Floquet symmetry-breaking (SB) phase and the other is a Floquet symmetry-protected topological (FSPT) phase, which we now discuss in order. [This is a good place to alert the reader to the literature on interacting Type I Floquet drives
and Floquet engineering, e.g. \cite{Grushin2014,Iadecola2015a,Klinovaja2016,Lee2018}, to which we will not do justice here.]

%%%%%%%%%%%%%%%%%%%%%%%%%%%%%%%%%%%%%%%%%%%%%%%%%%%%

%%%%%%%%%%%%%%%%%%%%%%%%%%%%%%%%%%%%%%%%%%%%%%%%%%%%
\subsection{Broken Symmetry Phases {\it aka} Time Crystals}
We first consider phases based on 
symmetry breaking. The first of these to be discovered was named the $\pi$-spin glass \cite{Khemani2016} on account of this symmetry breaking, but as these phases also break time translation symmetry, they have generally come to be known as Floquet time crystals \cite{Else2016b} or discrete time crystals \cite{YaoN2017,YaoN2017-E}. As this connection has a life of its own which would require a considerable digression, we will only touch upon it here; we direct readers to a review involving one of us for more details \cite{KMS-TCR}.

Returning to Fig.~\ref{classDPhasediagram} let us focus on the region labelled $\pi-$spin glass/time crystal. In this phase, all eigenstates exhibit Ising symmetry breaking but with a surprising twist. To see this
let us work along the line $t_0 = \pi/2$ and $0<t_1 < \frac{\pi}{2}$ where 
\beq
U(T) = e^{-  i t_1 H_1} \prod_{s} X_s \propto e^{-  i t_1 H_1} P \ .
\eeq
Thus, $U(T)$ has the form of an evolution with $H_1$ combined with a global spin flip in the $z$ basis. Readers can readily check that the Floquet eigenstates have the form of Ising symmetry breaking ``cat'' states
$$
| \pm \rangle = {1 \over \sqrt{2}} (| z_1,z_2,.....,z_n \rangle \pm | -z_1,-z_2,.....,-z_n \rangle),
$$
with eigenvalues $\pm e^{i \sum_s J_s Z_s Z_{s+1}}$. So unlike in the case of static symmetry breaking, there is 
now spectral pairing between states split by quasienergy $\pi/T$---whence the term $\pi$-spin glass. The term ``time crystal'' is intuitively justified by noting that a given starting state $| z_1,z_2,.....,z_n \rangle$ will only repeat after two periods. Really, there is order in both space and time and hence the phenomenon is properly described as spatio-temporal order \cite{vonKeyserlingk2016b}. Now this may seem non-generic to the reader on account of the special choice of starting state, but this is not the case---once interactions and randomness are fully taken into account, the late time states for \emph{generic} initial conditions also exhibit this period doubling. Even better, the starting microscopic Ising symmetry turns out to be unnecessary. The phase is {\it absolutely stable} in that any weak perturbation that preserves the period leaves the basic physics in place \cite{vonKeyserlingk2016b}. This stability can be attributed in part to the $\pi/T$ spectral gap between the paired eigenstates \cite{Else2016b}. The system generates an emergent Ising symmetry and a corresponding spatio-temporal order. Indeed, this makes the phase particularly easy to observe as it does not require fine tuning of either the drive or the initial state!

The analysis above has been extended to MBL Floquet drives with a completely broken on-site finite unitary symmetry group $G$. In this case the distinct Floquet drives are in correspondence with the elements $Z(G)$ of the center of the group; we direct the reader to \cite{vonKeyserlingk2016} for details. Finally we note that time crystalline period doubling also shows up in the clean, driven $O(N)$model at infinite $N$ \cite{Chandran2016} which has a very different, locally unbounded, Hilbert space although the model will heat once $N < \infty$.

%%%%%%%%%%%%%%%%%%%%%%%%%%%%%%%%%%%%%%%%%%%%%%%%%%%%
\subsection{Floquet Symmetry-Protected Topological Phases} 

We now turn to Floquet topological phases that are \emph{protected} by symmetry, known as FSPTs. Like their equilibrium counterparts (SPTs), FSPTs exhibit topological properties which are only robust if we demand that the system retain a protecting symmetry. If perturbations are allowed that break this symmetry, then the system can be brought into a trivial phase. FSPTs may be thought of as interacting generalisations of FTIs, in the same way that SPTs may be thought of as interacting generalisations of TIs. However, FSPTs exhibit a far richer phase structure than FTIs, and the full extent of this space is still being explored. 

In static systems, SPT order was initially considered a property of a gapped ground state (see Ref.~\citenum{Senthil2015} for a review). However, in the presence of MBL, the ordering properties of a given SPT phase may be shared by all (or a large fraction) of the eigenstates of the Hamiltonian \cite{Chandran2014,Slagle2015,Potter2015,Parameswaran2018}. The resulting eigenstate order inherits a robustness due to the stability of MBL, and cannot be removed without the system undergoing a delocalization transition. In Floquet systems, the lack of a well-defined ground state means that eigenstate order in the presence of MBL is the only physically meaningful notion of FSPT order. In particular, we will find that FSPT evolutions `pump' SPT phases to the boundary of an open system, which then exhibit eigenstate order in the presence of MBL.

Following standard notation, a Hamiltonian is protected by a unitary symmetry group $G$ if it satisfies $V(g)H(t)V(g)^{-1}=H(t)$ 
for all $t$ and for all $g\in G$, where $V(g)$ is the global representation of the group element $g$. If $G$ contains antiunitary elements (corresponding to a time-reversal symmetry), then the Hamiltonian satisfies $V(g)H(t)V(g)^{-1}=H(T-t)$ for those elements instead.

A topological classification of gapped ground state SPTs has been obtained using a range of techniques, in particular group cohomology, which we shall see also plays a key role in the classification of FSPT phases. Two Hamiltonians with the same symmetries are said to be in distinct SPT phases if one cannot adiabatically transform one Hamiltonian to the other while maintaining a gap and keeping the Hamiltonian at intermediate points in the same symmetry class. Equivalently, the ground state of an SPT phase cannot be obtained by applying a local finite-depth symmetry-preserving unitary transformation on a trivial ground state. In a system with a boundary, a 1d SPT phase has a degenerate ground state multiplet which provides a nontrivial (projective) representation of the group at the edges. In this case of 1d FSPT phases \cite{vonKeyserlingk2016,Else2016,Potter2016,Roy2016}, one finds that there arises a set of anomalous modes at the two edges. These correspond to sets of eigenstates which are identical in the bulk, but which differ at the edges, and whose quasienergies differ from each other in some quantized multiples.  
We will study a simple example of a 1d FSPT phase with $Z_2$ symmetry below. 

Consider again the binary Floquet model described by Eq.~(\ref{eq:classDUf}), but now with the specific parameter choice $t_1=\pi/2$ and $J_s=1$ for all $s$ ($J_s$ can be chosen to be random, but the present choice allows the physics to be demonstrated most clearly). This corresponds to the top boundary of the phase diagram in Fig.~\ref{classDPhasediagram}. As before, the model has a global Ising $\zbb_2$ symmetry $P=\prod_s X_s$, and the $h_s$ are random variables. At the end of a cycle (i.e., at $t=\pi/2$), for a system without a boundary, the unitary evolution operator is  $e^{-iH_{0}t_0}$. Thus $H_0$, which is truly localized, is the Floquet Hamiltonian and there is an absence of heating even when the Floquet cycle is repeated a very large number of times. 

For a chain with open boundary conditions, however, with end points labeled $1$ and $N$, the unitary at the end of the cycle is $Z_1 Z_n e^{-iH_{0}t_0}$. The edge unitary $\hat{Y}$ at the left end is thus $Z_1$. The action of this unitary on a simple product state $ \bigotimes_s \ket{x_s}$ where $\ket{x_s}$ is an eigenstate of $X_s$ with eigenvalue $x_s = \pm 1$ is, apart from an overall phase factor, to transform $x_1, x_N$ to $-x_1, -x_N$. The action at the edges may be viewed as a pump which changes the group representation of the edge spin states. The edge unitary cannot be realized in any zero-dimensional system which preserves the symmetry since $Z$ does not commute with $X$. The eigenstates of the open-boundary system thus come in quartets of two pairs whose quasienergies differ by $\pi/T$ \cite{Khemani2016,vonKeyserlingk2016,Else2016,Potter2016,Roy2016}.

In higher dimensions, it is also possible to construct such Floquet evolutions which change the representations of the group on disconnected boundaries. A more nontrivial kind of action is also possible. A $d$-dimensional SPT phase is characterized by the property that its ground state can be obtained from a trivial ground state by the action of a symmetry-preserving local unitary, which nevertheless cannot be locally generated at any finite depth. Remarkably, one finds that in all  dimensions $d$, Floquet evolutions can be constructed which behave as the identity in the bulk, and whose edge unitary is a $(d-1)$-dimensional pump which on successive application pumps a trivial product edge state through the set of topological SPT ground states~\cite{Potter2017,Roy2017}. This provides (at least a partial) classification of Floquet SPT phases in $d$-dimensions, which is the same as the classification of static SPT phases in $d-1$ dimensions. 

Classifications of these drives have been obtained in two different ways. Firstly, it has been conjectured that the classification of FSPT phases in $d$ dimensions with symmetry $G$ is given by the cohomology group $\mathcal{H}^{d+1}\left[\tilde{G},{\rm U(1)}\right]$, where $\tilde{G}=\zbb\times G$ if $G$ is unitary, and $\tilde{G}=\zbb\rtimes G$ if $G$ is antiunitary \cite{Else2016,Potter2016}. Secondly, an enumeration of FSPT states by direct construction has been obtained using a set of cohomology models for arbitrary $d$ and unitary $G$ in Ref.~\citenum{Roy2017} which agrees with this classification (though in the same work, FSPT phases which lie outside the classification were also found). In the unitary case, the cohomology group separates into two factors,
\bequ
\mathcal{H}^{d+1}\left[\zbb\times{G},{\rm U(1)}\right]=\mathcal{H}^{d+1}\left[{G},{\rm U(1)}\right]\times \mathcal{H}^{d}\left[{G},{\rm U(1)}\right],
\eequ
which may be associated with the Floquet Hamiltonian and loop component of the drive, respectively \cite{Roy2017}. Since the edge unitaries behave like SPT pumps, the order in these drives can be detected by time-dependent operators localized at the edges in much the same way as for time crystals \cite{vonKeyserlingk2016b}.

%%%%%%%%%%%%%%%%%%%%%%%%%%%%%%%%%%%%%%%%%%%%%%%%%%%%

\subsection{Many-body Chiral Floquet Phases\label{sec:mb_chiral_phases}}
Finally, we discuss a set of many-body Floquet phases that do not require any symmetry (other than time translation symmetry).
In Sec.~\ref{sec:RLBL_model}, we described a 2D single-particle Floquet model which exhibits protected chiral edge modes at the boundary of an open system even if the Floquet bands are topologically trivial \cite{Rudner2013}. Somewhat remarkably, the RLBL model is robust to the addition of interactions \cite{Nathan2017,Harper2017}. Moreover, many-body extensions to the RLBL model can be defined which exhibit chiral transport of \emph{information} at a boundary (as opposed to charge) \cite{Po2016,Harper2017}. In this section, we introduce the prototypical `SWAP' (or `exchange') model, and show how this model defines a set of many-body 2D Floquet topological phases without symmetry that may be classified by their boundary behaviour.

To define the model \cite{Po2016,Harper2017}, we retain the bipartite square lattice of the RLBL model, but now assume that each site hosts a finite-dimensional local Hilbert space of dimension $p$. Instead of particle hops, the fundamental unit of the model is the SWAP operator $\hat{\chi}_{\bf r,r'}$, which interchanges the local states on sites ${\bf r}$ and ${\bf r'}$ and can be generated by a local Hamiltonian (see Refs.~\citenum{Po2016,Harper2017} for explicit expressions). We will first assume that each site contains a single spin-1/2 degree of freedom (or equivalently, that the system consists of hardcore bosons), so that $p=2$. The drive then consists of four steps, $\hat{U}=\hat{U}_4\hat{U}_3\hat{U}_2\hat{U}_1$, with each step a product of SWAP operations over disjoint pairs of sites, $\hat{U}_s=\prod \hat{\chi}_{{\bf r}_B+{\bf b}_s,{\bf r}_A}$. The relevant pairs of sites in each step are exactly the same as for the RLBL model. In this way, Fig.~\ref{fig:AFAI} may be reinterpreted as describing the action of the SWAP model, if the highlighted bonds in each step now refer to SWAP operations instead of particle hops.

To understand the overall behaviour of the SWAP model, we consider how it acts on an initial many-body product state: on-site states in the bulk are permuted but return to their original site, while on-site states at a boundary are translated chirally by one site around the edge of the system. Since this holds for any product state, the complete action of the drive decouples into the identity operator in the bulk and a 1D edge unitary $\hat{Y}$ that acts only in the vicinity of the boundary. 

In this ideal model, the edge unitary acts as a pure 1D translation or `shift', which uniformly translates a many-body state around the periodic 1D boundary lattice. An operator of this kind is anomalous in that it is local, but cannot be generated by any local 1D Hamiltonian in a finite time \cite{GNVW,Harper2017,Po2016}; its eigenvalues are also nonlocal, and cannot be localised by any strength of disorder at the boundary. This property means that the translation is robust to perturbations that are generated by local 1D Hamiltonians acting only near the boundary. In turn, this motivates a classification of 2D Floquet drives based on their boundary behaviour: If we can classify all distinct classes of edge unitaries, equivalent up to 1D perturbations, then we can infer a corresponding classification of the bulk evolutions. 

\begin{figure*}
\center
\includegraphics[scale=0.5]{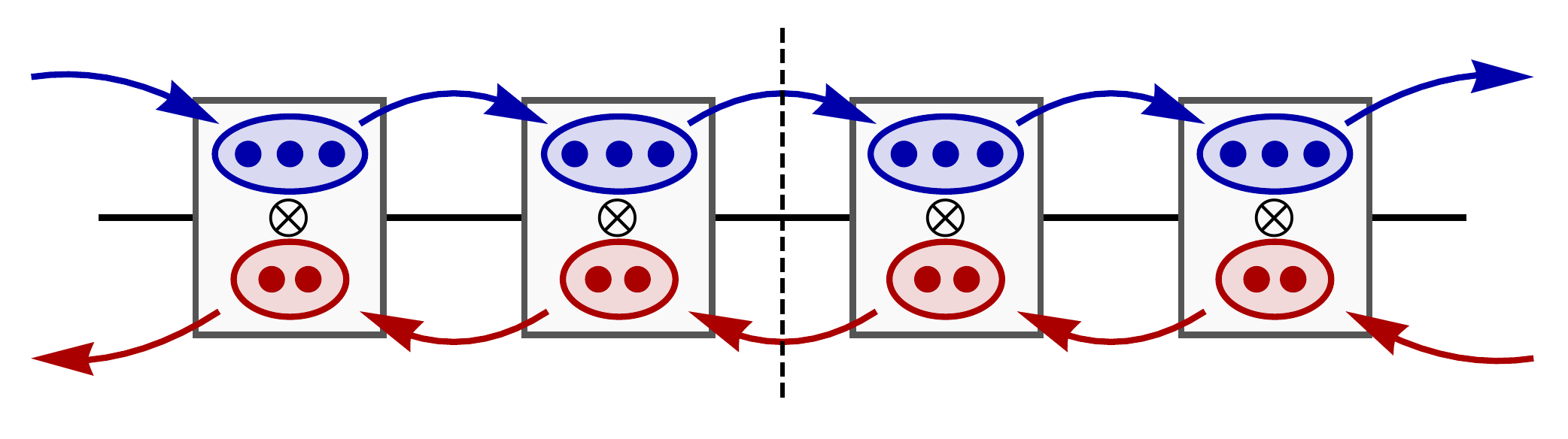}
\caption{A 2D chiral Floquet phase exhibits information flow at a boundary characterized by two coprime integers, $p$ and $q$. These indicate the size of the effective Hilbert space translated to the right or left (respectively) by the action of the drive, up to perturbations at the edge. The figure illustrates this action for $p=3$ and $q=2$, for on-site Hilbert spaces with total dimension 6. The chiral unitary index $\nu=\log(p/q)$ could be calculated directly by considering transport across the vertical dashed line, as described in Ref.~\citenum{Po2016}.  \label{fig:chiral_edge_pq}}
\end{figure*}

The SWAP model above resulted in the translation of each dimension $p=2$ Hilbert space by one site around the boundary. However, we could define similar drives for any finite on-site Hilbert space dimension $p$. More generally, we could run SWAP drives multiple times, or construct tensor products of drives with different values of $p$. A complete classification should determine which of these combinations have equivalent edge behaviours.
It may be verified that translating a $p=2$ Hilbert space twice to the right is equivalent to translating a $p=4$ space once to the right. Furthermore, if one has a stack of two $p=2$ drives, one moving to the left and one to the right by one site, the net drive is trivial and can be generated by a local 1D Hamiltonian. From considerations of this type, one can deduce that distinct drives are characterized by two coprime integers, $p$ and $q$~\cite{Harper2017}. These characterize the effective sizes of the Hilbert spaces moving to the left and to the right, illustrated schematically in Fig.~\ref{fig:chiral_edge_pq}.

An explicit index which can enumerate these different classes, and which confirms the intuitive picture described above, is the chiral unitary index $\nu(\hat{Y})$ (see Refs.~\citenum{GNVW,Po2016} for a complete definition of $\nu(\hat{Y})$, and methods for calculating it). 
Heuristically, the index calculates the degree to which local Hilbert spaces are transported chirally across an imaginary cut in the boundary by the action of $\hat{Y}$.
Importantly, it can be shown that $\nu(\hat{Y})$ is robust to locally generated unitary perturbations and always takes the discrete form $\nu=\log(p/q)$, where $p/q$ is a rational number \cite{GNVW,Po2016}, and therefore provides a well-defined topological invariant for the edge unitary of a 2D Floquet drive.  Ref.~\citenum{Po2016} interprets the index as quantifying the von Neumann entropy pumped along the boundary per Floquet cycle, discussed further in Ref.~\citenum{Duschatko2018}.

In this way, 2D unitary loop evolutions are partitioned into equivalence classes based on the information flow at their boundaries. As motivated in Sec.~\ref{sec:edge_unitaries}, an arbitrary many-body evolution often has a characteristic `loop component', and so these equivalence classes also provide a classification of 2D many-body Floquet evolutions more generally. In particular, if an evolution exhibits MBL at its end point, then the value of $\log(p/q)$ can be calculated directly \cite{Po2016}. It remains an open question whether a \emph{bulk} topological invariant for such an evolution can be found (analogous to Eq.~(\ref{eq:WindingNumber})). A related open question is the identification of a bulk quantity that can detect order in individual eigenstates. 
For the weakly interacting RLBL model, with its additional conserved quantity, orbital magnetization density may play such a role [see Refs.~\cite{Nathan2017a, Nathan2019}].

%%%%%%%%%%%%%%%%%%%%%%%%%%%%%%%%%%%%%%%%%%%%%%%%%%%%

%%%%%%%%%%%%%%%%%%%%%%%%%%%%%%%%%%%%%%%%%%%%%%%%%
\section{Experimental Manifestations}

Periodic driving has been used to investigate topological phenomena in a variety of experimental settings.
For example, topologically nontrivial Floquet bands have been created for photonic~\cite{Rechtsman2013, Hu2015,Mukherjee2017, Maczewsky2017, Mukherjee2018} and phononic~\cite{Peng2016}  systems, and for cold atoms in modulated optical lattices~\cite{Jotzu2014, Flaschner2016, Aidelsburger2015}.
Moreover, Floquet-MBL has been observed for cold atoms~\cite{Bordia2017}.
In solid state, topological Floquet gap opening on a topological insulator surface~\cite{Wang2013} and signatures of a light-induced quantized Hall effect in graphene have been observed~\cite{Oka2009, Gu2011, FoaTorresMultiTerminal, McIver2018}.

The above mentioned experiments demonstrate that many of the key ingredients for experimentally investigating the more exotic intrinsically dynamical (type II) many-body phases have been achieved.
In general we expect such systems to exhibit robust signatures of their nontrivial topology both through their bulk and edge properties.
For example, in the presence of disorder that localizes all bulk states, the RLBL system in a two-terminal transport setup hosts a quantized current at large source-drain bias~\cite{Kundu2017}; in a geometry without edges, the nontrivial topology of the system is manifested in a quantized average magnetization density for random many-body states~\cite{Nathan2017}. 
Similarly, the many-body swap model is predicted to host a quantized information flow on its edge~\cite{Po2016}; a corresponding readily measurable quantity remains to be identified, though ``in principle'' experiments have been proposed~\cite{Duschatko2018}. 

The most striking experimental progress directly germane to this review has come for broken symmetry phases/time crystals in a pair of twinned papers by the groups of Monroe \cite{Zhang2017} and Lukin \cite{Choi2017}. The first experiment realized a variant of our model binary Ising drive wherein the bonds are non-random and instead disorder is introduced via a random longitudinal field. While the resulting model does not have a time crystal phase in the infinite system limit \cite{KMS-TCR}, the relatively small ion trap system that was studied exhibited a striking enhancement of period doubled oscillations starting from a fully polarized state. The second experiment studied a much bigger system of NV centers which realizes another effective binary Ising drive but this time with random exchange interactions with a dipolar decay in space. The experiment again observed systematically enhanced period-doubled oscillations starting from a polarized state. This system again does not truly realize the ideal time crystal as MBL is not stable for interactions of this range. In subsequent work \cite{Ho2017} it was argued that this system exhibits a critical time crystal, in which correlations have an algebraic decay in time. 
We note that FSPTs, at least in $d=1$ can potentially be detected by a similar period doubling for edge observables \cite{vonKeyserlingk2016b}.

\section{Recent Progress}
%%%%%%%%%%%%%%%%%%%%%%%%%%%%%%%%%%%%%%%%%%%%%%%%%%%%
%
Finally, we provide a brief description of some of the other topics under active investigation in this field. In the single-particle case, a number of works have derived explicit bulk-edge correspondences (and real-space topological invariants) for FTI phases in the presence of disorder \cite{Asboth2013,Graf2018,Liu2018}. Bulk-edge correspondences for 1D systems have also been obtained in the context of quantum walks \cite{Cedzich2016} and more generally using K-theory \cite{Sadel2017}. Beyond this, a Wannier representation of FTI states was introduced in Ref.~\citenum{Nakagawa2019}, while the possibility of realizing so-called \emph{higher-order} FTI phases was studied in Refs.~\citenum{Huang2018,Bomantara2019,Rodriguez-Vega2018}. In a similar direction, several groups have also studied the possibility of Floquet topological phases protected by nonsymmorphic `space-time crystalline' symmetries \cite{Morimoto2017,Xu2018,Peng2018}.

In the many-body case, the SWAP model described in Sec.~\ref{sec:mb_chiral_phases}, and the associated 2D Floquet chiral phases, have inspired several other exotic many-body phases. An extension of the SWAP drive to 3D was studied in Ref.~\citenum{Reiss2018}, while in a pair of papers \cite{Fidkowski2019,Po2017}, the SWAP model was extended to 2D fermionic systems, which it was shown may host \emph{radical} chiral phases with a dynamical $\mathbb{Z}_2$ index. In Ref.~\citenum{Po2017}, Radical Floquet phases were also shown to arise in a driven version of Kitaev's honeycomb model \cite{Kitaev2006}, with evolutions in this class argued to exhibit `Floquet Enriched Topological' order. This has natural extensions to other Abelian and non-Abelian groups, studied in detail in Ref.~\citenum{Potter2017}.

Floquet chiral phases also arise in the construction of FSPT phases which lie beyond the cohomology construction. In Ref.~\citenum{Roy2017}, it was demonstrated that chiral Floquet order may coexist with FSPT order. The same reference also showed that counterpropagating chiral drives corresponding to two factors of an internal symmetry group, $G=G_1\times G_2$, which have no \emph{net} transfer of information at the boundary, are not necessarily equivalent to a trivial drive in the presence of symmetry. A set of beyond-cohomology FSPT phases protected by time-reversal symmetry, dubbed Floquet topological paramagnets, was also introduced in Ref.~\citenum{Potter2018}. The nature of the transition between different many-body Floquet phases in general remains an open question. However, significant progress towards understanding the transitions in the model of Eq.~(\ref{eq:classDUf}), and the associated notion of Floquet quantum criticality, was made in Ref.~\citenum{Berdanier2018}. Work in this area is ongoing.

%%%%%%%%%%%%%%%%%%%%%%%%%%%%%%%%%%%%%%%%%%%%%%%%%%%%
\section{Outlook}
%%%%%%%%%%%%%%%%%%%%%%%%%%%%%%%%%%%%%%%%%%%%%%%%%%%%
Periodically driven systems offer a fascinating arena in which to study physics that is qualitatively different from that of static systems. Lying at the confluence of several distinct fields, a full appreciation of their properties requires concepts from topology, symmetry-breaking, quantum information, and localization physics. Perhaps most excitingly, technical advances in the area of cold atoms suggest that many of these exotic phases could well lie within experimental reach. In this review, we have tried to give a taste of the key concepts and types of phases that may arise in Floquet systems, as well as arguments for how they may be stabilized by disorder and realized experimentally. We close by discussing some of the outstanding challenges that remain in this rapidly advancing field. 

At this point the full set of theoretical ideas is in advance of experimental realizations and so the most desirable progress would consist of finding and studying such realizations. Synergistically, it would be good to have simpler theoretical proposals for the detection of two of our three sets of many-body phases, the FSPTs and the chiral phases. While we primarily have cold atomic systems in mind, the study of solid state systems is more complex but necessary and potentially quite rewarding. More theoretically, gaining an understanding of the bulk invariants associated with various topological Floquet phases is an open problem and also points to the project of obtaining a complete classification of such phases. Finally, MBL is very likely not truly stable in $d>1$, and a detailed understanding of its limitations in the Floquet setting and of various pre-thermal regimes will help us understand to what extent the higher dimensional phases suggested by the first round of theory are ``good enough for government work''. We invite the reader to make progress on any and all of these questions!

\begin{acknowledgments}
We thank C. von Keyserlingk, V. Khemani, A. Lazarides and W. Ho for useful comments on the manuscript. We also gratefully acknowledge support from the NSF under CAREER DMR-1455368 (F.H. and R.R.); the European Research Council (ERC) under the European Union Horizon 2020 Research and Innovation Programme (Grant Agreement No. 678862), and the Villum Foundation (M.S.R.); and the US Department of Energy grant No. DE-SC0016244 (S.L.S.). This research was also developed with funding from the Defense Advanced Research Projects Agency (DARPA) via the DRINQS program. The views, opinions and/or findings expressed are those of the authors and should not be interpreted as representing the official views or policies of the Department of Defense or the U.S. Government. 
\end{acknowledgments}

\end{document}